\documentclass[preprint]{aastex62}

\usepackage{amssymb}
\usepackage{gensymb}
\usepackage[english]{babel}
\usepackage{wasysym}

\hypersetup{linkcolor=red,citecolor=blue,filecolor=magenta,urlcolor=cyan}

\newcommand\ie{{\it i.e., }}
\newcommand\eg{{\it e.g., }}

\begin{document}
\title{From Centaurs to comets --- 40 years} 

\author[0000-0002-6830-476X]{Nuno Peixinho}
\correspondingauthor{Nuno Peixinho}
\email{peixinho@mat.uc.pt}
\affiliation{CITEUC -- Center for Earth and Space Research of the University of Coimbra, 
Geophysical and Astronomical Observatory of the University of Coimbra, 3040-004 Coimbra, Portugal}

\author[0000-0002-1506-4248]{Audrey Thirouin}
\affiliation{Lowell Observatory, 1400 W Mars Hill Rd, Flagstaff, AZ 86001, USA.}

\author[0000-0000-0000-0000]{Stephen C. Tegler}
\affiliation{Department of Physics and Astronomy, Northern Arizona University, Flagstaff, AZ 86011, USA}

\author[0000-0000-0000-0000]{Romina P. Di Sisto}
\affiliation{Instituto de Astrof\'{\i}sica de La Plata, CCT - CONICET - Universidad Nacional de La Plata, Argentina}
\affiliation{Facultad de Ciencias Astron\'{o}micas y Geof\'{\i}sicas, UNLP, La Plata, Argentina}

\author[0000-0000-0000-0000]{Audrey Delsanti}
\affiliation{Aix Marseille Universit\'{e}, CNRS, LAM (Laboratoire d'Astrophysique de Marseille) UMR 7326, 13388 Marseille, France}

\author[0000-0000-0000-0000]{Aur\'{e}lie Gilbert-Lepoutre}
\affiliation{LGL-TPE, UMR 5276 / CNRS, Universit\'{e} de Lyon, Universit\'{e} Claude Bernard Lyon 1, ENS Lyon, 69622 Villeurbanne Cedex, France}

\author[0000-0000-0000-0000]{James G. Bauer}
\affiliation{University of Maryland, Department of Astronomy, College Park, MD 20742-2421, USA}

\begin{abstract}
In 1977, while Apple II and Atari computers were being sold, a tiny dot was observed in an inconvenient orbit. The minor body 1977~UB, to be named (2060) Chiron, with an orbit between Saturn and Uranus, became the first Centaur, a new class of minor bodies orbiting roughly between Jupiter and Neptune. The observed overabundance of short-period comets lead to the downfall of the Oort Cloud as exclusive source of comets and to the rise of the need for a Trans-Neptunian comet belt. Centaurs were rapidly seen as the transition phase between Kuiper Belt Objects (KBOs), also known as Trans-Neptunian Objects (TNOs) and the Jupiter-Family Comets (JFCs). Since then, a lot more has been discovered about Centaurs: they can have cometary activity and outbursts, satellites, and even rings. Over the past four decades since the discovery of the first Centaur, rotation periods, surface colors, reflectivity spectra and albedos have been measured and analyzed. However, despite such a large number of studies and complementary techniques, the Centaur population remains a mystery as they are in so many ways different from the TNOs and even more so from the JFCs.  
\end{abstract}

\section{Introduction}
\label{sec:intro}

Centaurs are a group of Solar System minor bodies loosely classified as those orbiting around the Sun between the orbits of Jupiter and Neptune that are not trapped in any mean-motion resonance with a planet. 
They have been discovered 15 years before the Trans-Neptunian Objects (TNOs), or Kuiper Belt Objects (KBOs), except for Pluto/Charon. 
First announced as ``slow-moving object Kowal", on November 4$^{th}$, 1977, and then with the provisional designation 1977~UB, on November 14$^{th}$, when it was shown that a very high eccentricity orbit was not viable for it, (2060) Chiron was the first Centaur to be discovered \citep{1977IAUC.3129....1K}. Chiron was also precovered in photographic plates from 1941 and even from 1895 at the Boyden Observatory, South Africa \citep{2001bepl.book.....D}.

\citet{Fernandez1980} suggested that Chiron might be a comet 
diffusing inwards, through planetary perturbations, from a cometary belt beyond Neptune---the by then undetected trans-Neptunian belt. But the orbit of Chiron was not its only unusual characteristic. \citet{1988IAUC.4554....2T} found that its brightness increased more than expected 
and, later on, \citet{1990AJ....100.1323M} directly detected a low-surface-brightness coma for Chiron. Then, it was hypothesized that Chiron was potentially an ice-rich object, a comet nucleus 180~km across, darkened by C-class carbonaceous soil, that has been scattered inward from the Oort cloud \citep{1990Icar...83....1H}. A second idea was that Chiron is most likely a bright member of what is known today as the trans-Neptunian belt \citep{1988ApJ...328L..69D}.
Finally, \cite{1990Natur.348..132H}, concluded that Chiron-type orbits are short-lived and chaotic, with an ejection half-life of $t_{1/2}\sim1$~Myr, therefore Chiron's current orbit should be a transition between a trans-Neptunian object and a short-period comet. 

In 1992, seven months before the discovery of the first TNO, a second Centaur was announced: 1992~AD, now 
known as (5145) Pholus \citep{1992IAUC.5434....1S}. No sign of coma was identified, but Pholus became the reddest known object in the Solar System \citep{1992IAUC.5450....2H}. 

Given their close link, Centaurs have been mostly studied together with TNOs. 
Previous to this book, the most extensive review of this field was the book ``The Solar System Beyond Neptune''  \citep{2008ssbn.book.....B},  
preceded by works as the ``Trans-Neptunian Objects and Comets'' \citep{2008tnoc.book.....J} and the pioneering  
``The first decadal review of the Edgeworth-Kuiper Belt'' \citep{2004fdre.book.....D}. Other reviews, such as
\cite{2018ARA&A..56..137N}, \cite{2018SSRv..214..111F}, \cite{2002ARA&A..40...63L}, and \cite{1999AREPS..27..287J} are also of highly recommended reading. 
In this chapter, we will propose a brief overview of the current state of the art 
regarding composition, rotational and physical properties, and activity regarding the Centaur population 
as well as their links to the trans-Neptunian and comet populations. 


\section {Centaurs as progeny of TNOs}
\label{sec:progenTNO}

\cite{1997Icar..127...13L} and \cite{1997Sci...276.1670D} found that a belt of TNOs on rather circular and low inclined 
orbits---presently known as: Classical TNOs--- 
would create a dynamically excited population through scattering interactions with Neptune, hence called the Scattered Disk Objects (SDOs), 
and that population should be the main source of Centaurs. Integrating different models for that scattered population for $4.5$ Gyr 
provides injection rates of $\sim(1-5)\times10^{-10}$ SDOs$/$yr and an estimated 
number of Centaurs larger than $\sim 1$ km of the order of $10^6$-$10^8$
\citep{2007Icar..190..224D, 2008ApJ...687..714V}. Plutinos, the TNOs on the 3$:$2 mean motion resonance (MMR) with Neptune, 
should also be a source of Centaurs \citep{1997Icar..127....1M, 1999AJ....118.1873Y}, contributing $\sim 3$ times less 
than SDOs \citep{2007Icar..190..224D}, and the 2$:$1 resonant TNOs might also contribute \citep{2009AJ....138..827T, 2008A&A...490..835D}. Although very secondary as source, every $\sim  60 - 200$ yr a kilometer-sized Neptune Trojan enters the Centaurs 
region \citep{2010MNRAS.402...13H}. Jupiter Trojans were also found to be a minor source of Centaurs \citep{2019Icar..319..828D}.

The mean lifetime of Centaurs is strongly dependent on their definition, because lifetime correlates with 
perihelion distance and semi-major axis and with the initial inclination of the 
progenitors, and can be as short as few Myr up to tens of Myr 
\citep{2003AJ....126.3122T, 2007Icar..190..224D, 2008ApJ...687..714V, 2009Icar..203..155B}. 

Interesting is the very low probability for a high inclination Centaur, or even a retrograde one, 
originating from the trans-Neptunian region \citep{2013Icar..224...66V} and 
a heterogeneous group of up to 200 Centaurs with diameters $D \gtrsim 150$ km, inclinations $i > 70^{\circ}$, and perihelia $q >15$ au should exist, originating mainly in the Oort Cloud \citep{2012MNRAS.420.3396B, 2014Ap&SS.352..409D}.

Another very interesting aspect of Centaurs is the existence of binaries. From the 81 known trans-Neptunian binaries (TNBs), 
(42355) Typhon/Echidna and (65489) Ceto/Phorcys cross the orbit of Neptune. 
\cite{2014MNRAS.437.2297B} conclude that a small population of contact binary Centaurs may exist and \cite{2018MNRAS.476.5323A} find that Typhon/Echidna could survive close encounters with the planets and even reach the terrestrial planets region as a binary. 


\section {Centaurs as progenitors of JFCs}
\label{sec:progenJFC}

On early days, the expression short-period comets (P$<$200~yr) was frequently used and the trans-Neptunian belt was thought to be their main source. Centaurs were merely a transient phase. The usage of short-period comets was replaced by the more concrete Jupiter-family comets (JFCs) 
and Halley-type comets (HTCs)---Chiron-type and Encke-type are also used---, using not the orbital period as a criterion but the Tisserand parameter relative to Jupiter, $T_J$ \citep{Tisserand1896, Kres1972, 1996ASPC..107..173L}. JFCs are those with 
$2\lesssim T_J \lesssim 3$. 

It is interesting to notice that the studies regarding the source of JFCs have been made, mostly, regarding the direct link between them and TNOs. Centaurs are taken as a short-lived transient phase not decoupled from this JFC-TNO link.  
Right after the discovery of Pluto, \cite{1930ASPL....1..121L} speculated about the existence of trans-Neptunian and trans-Plutonian objects but 
without linking them explicitly to comets. The existence of a trans-Neptunian region as a source of JFCs was first 
suggested by \cite{1943JBAA...53..181E,1949MNRAS.109..600E} and \cite{1951PNAS...37....1K}, being properly addressed much later by \citet{Fernandez1980} and \citet{1988ApJ...328L..69D}. 

Relatively little attention has been given to the dynamics of the Centaur population itself.
\cite{2003AJ....126.3122T} find that one-third of the Centaurs will actually be injected into the JFC population, while two-thirds 
will be ejected from the Solar System or will enter the Oort cloud. About 20\% of them have lifetimes as Centaurs shorter than 1 Myr, while another 20\% have lifetimes exceeding 100 Myr. \citet{2004MNRAS.355..321H} found even that about one-fifth of the Centaurs that became JFCs will become Earth-crossing objects. Some Centaurs may even become temporarily captured as Trojans of the giant planets 
for up to 100~kyr \citep{2006MNRAS.367L..20H}. Most interesting is the recent finding that active Centaurs tend to evolve to JFCs while the inactive ones tend to evolve to Halley-type comets \citep{2018P&SS..158....6F}. 


\section {Centaurs by themselves}
\label{sec:themselves}

\subsection{Classifying Centaurs}

\begin{figure*}
\includegraphics[width=18cm, angle=0]{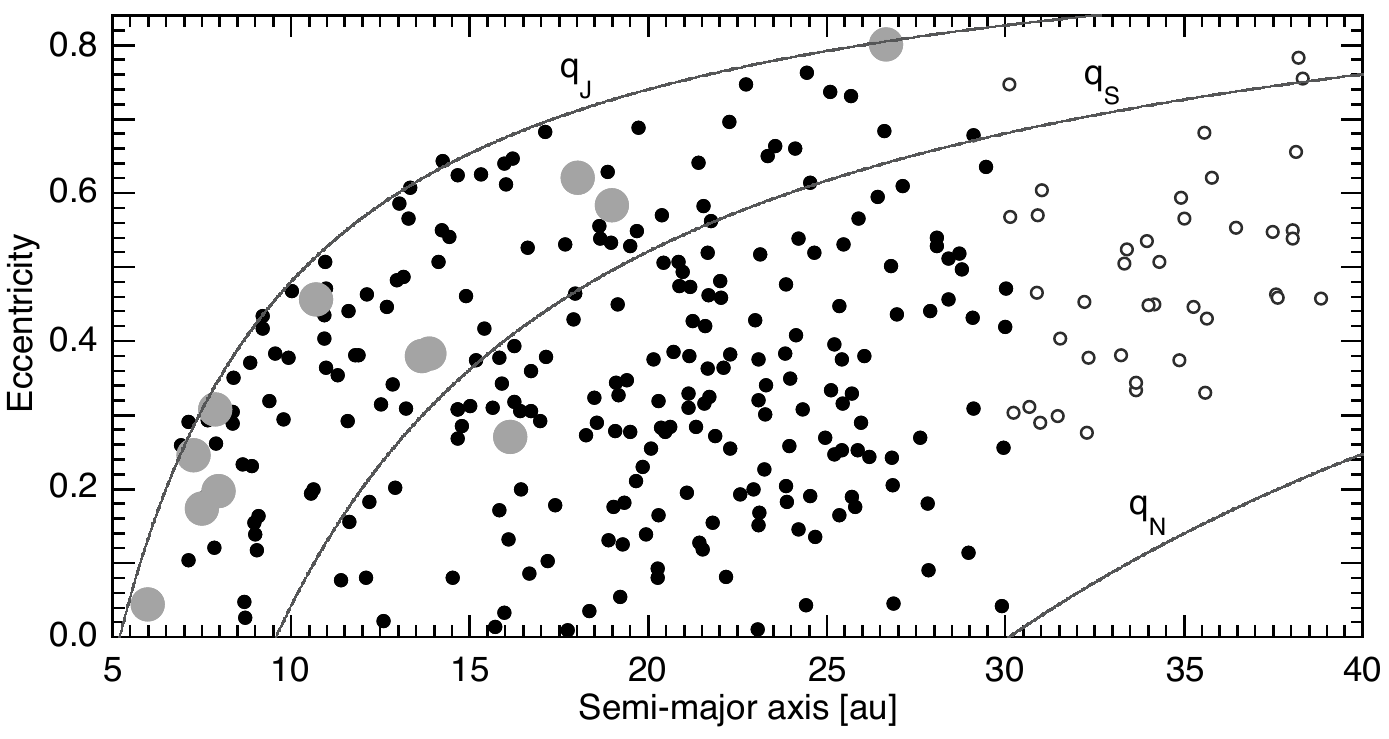}
 \caption{Semi-major axis and eccentricity plot of known Centaurs. Large gray dots indicate known active Centaurs. 
 Black dots indicate Centaurs with both perihelion and semi-major axis 
 between those of Jupiter and Neptune, \ie the more strict definition. White dots indicate objects classified 
 as Centaurs by DES alone. The curves where objects have perihelion equal to the semi-major axis of 
 Jupiter, $q_J$, of Saturn, $q_S$, and of Neptune, $q_N$, are drawn. Several objects are out of scale.}
\label{fig:a_e}
\end{figure*}

There is no strict definition of Centaurs, which poses some evident problems when comparing different works. 
The most frequent classification forces them to have 
both perihelion and semi-major axis between the orbits of Jupiter and Neptune
($q>5.2$~AU, $a<30.1$~AU) \citep[\eg ][]{2008ssbn.book...43G}. With the previous definition, the most strict, 
as of February 1$^{st}$ of 2019\footnote{\url{https://ssd.jpl.nasa.gov/sbdb_query.cgi}} there are 292 Centaurs, 
of which 26 are also classified as comets.  
Also widely used is the Deep Ecliptic Survey (DES) definition, in which the perihelion needs to be between the orbits of Jupiter and Neptune but without any constraints regarding the semi-major axis \citep{2005AJ....129.1117E}. With this definition, there are 360 Centaurs listed\footnote{\url{http://www.boulder.swri.edu/~buie/kbo/desclass.html}}. See Figure \ref{fig:a_e}.


\subsection{Surface properties}

\subsubsection{Colors}
\label{sec:colors}

When photometry is acquired almost simultaneously in various bands, or the eventual 
rotational brightness variations are negligible or averaged down, the magnitude difference difference between those bands 
defines a color. Centaurs' colors are usually measured using the Johnson UBVRJHK photometric system---from $\approx 3600$ {\AA} to $\approx 2.2 \mu$m---, or the SDSS $ugriz$ system \citep[\eg ][]{Schwamb2018}. 
In the most used color, Centaurs range from a neutral/gray solar-like B$-$R $=$ 1.0~mag to an 
extraordinarily red B$-$R $=$ 2.0~mag. Since complex organic molecules are known to efficiently absorb shorter 
wavelengths of optical light, the redder the object the richer the object might be in complex organic molecules, 
although other compounds are capable of doing so. Surface colors can also be used to obtain the low resolution reflectance spectrum, 
at a relatively low observation time cost and constraint, at least, the spectral continuum of the studied object
\citep[see][]{2004A&A...417.1145D, 2015Icar..252..311D}. 
Two large surveys post their measurements of colors for these objects 
online the Tegler, Romanishin, and Consolmagno's survey, TRC\footnote{\url{http://www.physics.nau.edu/~tegler/research/survey.htm}}, 
and the Minor Bodies of the Outer Solar System survey, MBOSS\footnote{\url{http://www.eso.org/~ohainaut/MBOSS}}. 

Early on, in the observations of Centaurs and TNOs, a controversy arose as to whether these objects exhibited a unimodal color distribution \citep{1996AJ....112.2310L, 2001AJ....122.2099J, 2002A&A...389..641H} or a bimodal color distribution 
\citep{1998Natur.392...49T, 2003Icar..161..181T, 2003ApJ...599L..49T}. 
Models attempting to explain the two possible behaviours were very different. 
\cite{1996AJ....112.2310L} put forth a model of steady radiation reddening and stochastic impacts to explain the unimodal colors of Centaurs and TNOs, but it has been shown that such simple model could not explain the color diversity as it was observed 
\citep{2001AJ....122.2099J, 2003Icar..162...27T, 2003EM&P...92..233T}. 
Posterior modelling and laboratory works argue that it is, in fact, possible to reproduce the whole range of colors observed 
with an appropriate combination of initial albedo, meteoritic bombardment, and space weathering \citep{2012Icar..221...12K}.
On the other hand, \cite{2003ApJ...599L..49T} argue in favor of a thermally driven composition gradient in the primordial trans-Neptunian belt that 
would result in bimodal colors. \cite{2011ApJ...739L..60B}  and \cite{2016AJ....152...90W} argue in favor of H$_2$S, rather than simple organic molecules, as the main reddening agent, where objects located beyond the H$_2$S sublimation line, hence not depleted of it, would suffer 
radiation reddening contrarily to the depleted ones that would result in gray surfaces. 

\begin{figure*}
\includegraphics[width=18cm, angle=0]{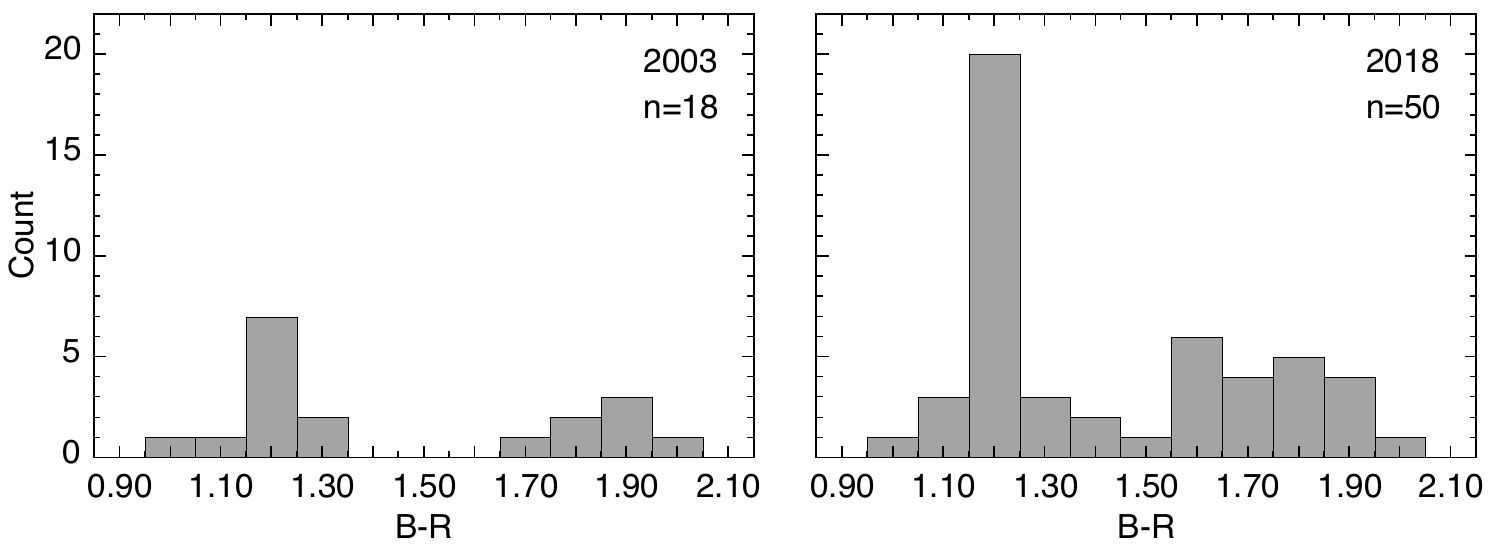}
 \caption{Centaur B$-$R histograms as they were in 2003 and as they are in 2018 from TRC data. 
In the larger 2018 sample the red population became broader and the gap between the gray and red populations 
became a dip. Note that we are using the more strict definition of Centaurs. }
\label{fig:histograms}
\end{figure*}

We now see that Centaurs and different dynamical classes of TNOs, as well as different sizes, exhibit different colors
\citep[\eg][]{2008ssbn.book..105T, 2012A&A...546A..86P, 2012ApJ...749...33F, 2016AJ....152..210T, 2017AJ....153..145W}.
Consensus is growing towards a scenario where migration of outer planets disrupted the primordial disk of icy planetesimals, 
objects were scattered onto dynamically-hot orbits and then onto Centaur orbits to present us with the bimodal color we see today. 
In contrast, dynamically-cold TNOs were far enough away from the migrating outer planets that they were not scattered onto 
dynamically-hot TNO orbits and so represent a third compositional class. 
Nonetheless, recent observations call the bimodality of Centaurs and, therefore, models to explain the bimodality into question. 
The 99.5\% confidence level for the bimodality found by \cite{2003A&A...410L..29P} and \citep{2008ssbn.book..105T} 
drops to 81\% with today's larger sample \citep[][using TRC survey data]{2016AJ....152..210T}. 
Even reanalyzing TRC data for the 50 objects that follow \citeauthor{2008ssbn.book...43G}'s more strict 
definition of a Centaur, we get a confidence level of 91.6\%, \ie no different conclusion. Looking at Figure \ref{fig:histograms} 
we can see how the red population has a larger spreading today. 
It is interesting that \cite{2011ApJ...739L..60B} predicted that the red population should have a broader color distribution than the gray one, given the several reddening agents with sublimation lines in the 20 to 35 au region of the primordial trans-Neptunian belt---although in their most recent paper, they claim only one species, H$_2$S, is serving as the reddening agent. 
It is, evidently, premature to rule out irradiation of simple organic molecules into complex organic molecules in favor of irradiation H$_2$S without further evidence, and without studying what would happen when ices are not in their pure form but in a certainly more realistic mixture with refractory material and how would that affect the sublimation lines. 

Further B$-$R observations of Centaurs are essential. Confirmation of bimodal Centaur colors will bolster models like that of \cite{2016AJ....152...90W}. On the other hand, refuting bimodal colors for Centaurs will require a rethinking of the important processes acting on Centaurs. The fact that red Centaurs have a 
smaller orbital inclination angle distribution than the neutral ones \citep{2016AJ....152..210T}, suggesting a distinct origin, 
is also puzzling when dynamical models say that Centaurs do not preserve their inclinations \citep{2013Icar..224...66V}. 
Moreover, the recent discovery that Jupiter Trojans and Neptune Trojans 
possess similar colors, when they were presumably populated by very distinct regions of the primordial planetesimal disk, 
evidences that we are still missing something crucial \citep{2018AJ....155...56J}.  

\subsubsection{Spectra}

Although surface colors provide a good proxy to the general surface properties of our objects, 
the dedicated investigation tool for a comprehensive surface composition study is, of course, reflectance spectroscopy. 
The technical limit to this method is set by the collecting power of the current large telescopes and the sensitivity of the instruments. 
With an 8-10 m class telescope, visible spectroscopy with a signal-to-noise ratio (SNR) above $\sim$20 is possible only for objects with V$<$20-22. Near-IR spectroscopy can be performed at SNR $\gtrsim$10 for objects with J$<$18-20. Therefore, only a subset of the currently known Centaurs can be explored with this technique. 

Visible reflectance spectra ($\lambda \sim 400-700$ nm) generally show a linear increasing continuum.
Most of the ices expected to be present in these regions of the Solar System have 
absorption bands in the near-infrared (NIR). Signatures of water ice, H$_2$O, are present at 1.5, 1.65, and 2.0 $\mu$m;  
of methane ice, CH$_4$, at 1.7 and 2.3 $\mu$m; of methanol ice, CH$_3$OH, at 2.27$\mu$m; and of 
ammonia ice, NH$_3$, at 2.0 and 2.25 $\mu$m. 
Unfortunately, observations in the region $\lambda > 3 \mu$m, where ices of interest have very deep signatures 
(including also CO and CO$_2$), are not feasible. However, some predictions on the surface composition 
have been tempted by \cite{2015Icar..252..311D}. 

(5145) Pholus masterfully opened the era of the spectroscopic exploration of the outer Solar System belt 
and what we know today about other objects is just a briefer story.    
Since Pholus, about two dozens of centaurs were studied with spectrographs from the largest facilities available from ground. 
Its first spectrum showed to be linear between 400 nm and 1 $\mu$m, being the steepest red slope detected so far on a Solar System body 
\citep{1992Icar...97..145F, 1992Icar...99..238B, 1993JGR....98.7403H}. 
Such slope was incompatible with the conventional silicate or meteoritic analogs that prevailed in asteroids studies at that time, 
but rather compatible with a mixture of tholins \citep{1989Icar...79..350K}, suggesting the presence of organic material processed by irradiation. 
\cite{1992Icar...99..238B} suggested that this body has a primitive composition and is at either the beginning or the end state of its thermal evolution. Although on an unstable orbit, Pholus has been undisturbed for at least $10^5$ yr, \ie 
long enough to hold, possibly, a thin surface crust of organics due to space weathering. 
If Pholus just entered the Centaur region, it should start to experience heating and surface processing. 
In alternative, it might have experienced past cometary activity, maybe on a different orbit in that region, and rather be at its 
thermal end state with an insulating mantle of organic tholins. 
Using \cite{1993tres.book.....H} theory, laboratory data, and new observations,  \cite{1998Icar..135..389C} 
conclude that the red color of Pholus is caused by the irradiation processing of simple molecules 
(CH$_4$, H$_2$O, CO, NH$_3$, N$_2$, CH$_3$OH) and describe the full spectrum of Pholus with the presence of silicate olivine, 
carbon, a complex refractory solid (organic molecules), and a mixture of H$_2$O and CH$_3$OH (or an equivalent molecule), 
highlighting that this inventory includes all the basic components of comets. A nice display of Pholus NIR spectrum 
can be found in \cite{2011Icar..214..297B}. 

Spectroscopically, Centaurs seem to exhibit the same variety of features as their TNO progenitors, except for 
methane, seen only among the large TNOs. Water ice has been detected on 
(2060) Chiron, (8405) Asbolus, (10199) Chariklo, (31824) Elatus, (32532) Thereus, (52872) Okyrhoe, (55576) Amycus, and 2007~UM$_{126}$, 
and methanol has been detected on (5145) Pholus, (54598) Bienor, (83982) Crantor, and 2008~FC$_{76}$ (see Table \ref{Tab:LCs}).


\subsection{Rotations, Shapes and Sizes}


\subsubsection{Lightcurves}
\label{sec:rotation}

The rotational brightness variation, commonly named \textit{lightcurve}, is the periodic variation of an object brightness as a function of time due to its rotation. The lightcurve can be produced by various mechanisms: i) albedo variations on the body surface, ii) elongated shape, and/or iii) contact/close binary 
\citep[\eg ][]{2002AJ....124.1757S, 2005PhDT........21L}. Therefore, lightcurves carry a lot of information about the rotational and physical properties of small bodies.  

The first two basic parameters derived from lightcurve are the rotational period of the object (P), and the peak-to-peak lightcurve amplitude ($\Delta m$). 
Assuming a triaxial object in hydrostatic equilibrium, one can estimate the lower limit to its density using the figures of equilibrium from \citet{1987efe..book.....C}. Assuming the lightcurve is dominated by the shape of the object, 
one can estimate its elongation assuming a certain viewing angle. 
A lighturve can also be caused by a spherical object with albedo variation(s) on its surface and so one can extract constraints. 
Finally, a lightcurve with a U- and V-shape at the maximum and minimum of brightness and $\Delta m>0.9$~mag is likely due to a contact/close binary, and thus we can infer the binarity of the object and derive the physical characteristics of both system's components \citep{Leone1984, Lacerda2014SQ317, Thirouin2018}.

Only $\sim$5$\%$ of the known Centaurs have been observed for complete/partial lightcurves (Table~\ref{Tab:LCs}) showing, nonetheless, a large variety of amplitudes and rotational periods.
From Maxwellian fits the mean rotation period of the Centaurs is 8.1~h, being slower than the mean period of TNOs, \ie 8.45~h \citep{Thirouin2016, Thirouin2019}. The mean amplitude is $\sim0.15$~mag. Because of Centaurs relatively short orbital periods, 
compared to the TNOs, the viewing angle between the rotation axis and the line of sight will change significantly 
over the course of a couple of years, and thus the lightcurve amplitude will vary, as seen for Bienor and Pholos 
\citep{Tegler2005, FernandezValenzuela2017}. 
In such cases, it is possible to 
derive the object's pole orientation. 
So far, no strong correlations between 
P and $\Delta m$ and orbital parameters have 
been detected among Centaurs \citep{2009A&A...505.1283D}. 
But the correlation between lightcurve amplitude and absolute magnitude (smaller objects are more elongated than the larger ones) 
noticed among TNOs is not present in the Centaur population which infer for a different evolution/collisional history \citep{2009A&A...505.1283D, 2013PhDT.........T, 2013AJ....145..124B}.

 
\subsubsection{Size distribution}  

The radius of an object (in km) is given by: ${p_R}\times R^2 = 2.24\times10^{16}\times 10^{0.4\times(R_{\astrosun}-H_{R})}$, 
where R$_{\astrosun}$ is the R-filter magnitude of the Sun, p$_{R}$ and H$_{R}$ are, respectively, the geometric albedo and the absolute magnitude of the object in the R-filter \citet{1916ApJ....43..173R}. 

Several techniques can be used to derive or constrain the albedo and absolute magnitude and by extension the size of an object. Stellar occultations can be used to derive shape, size, albedo, detection of rings and satellites, and constraints about 
atmosphere, such as for Chiron and Chariklo \citep{2014Natur.508...72B, 2015Icar..252..271R, 2015A&A...576A..18O}.
With data from the Spitzer Space Telescope, the Herschel Space Observatory, the Wide-field Infrared Survey Explorer (WISE) mission, and the Atacama Large Millimeter Array (ALMA) it has been possible to perform thermal modeling in order to estimate the radiometric size and albedo \citep{2008ssbn.book..161S, 2013A&A...557A..60L, 2013ApJ...773...22B, 2014ApJ...793L...2L, 2014A&A...564A..92D,  2017A&A...608A..45L}. 
Generally, Centaurs have a low albedo, of a few percent, and only a handful have an albedo up to $\sim$20$\%$ \citep{2014A&A...564A..92D, 2014ApJ...793L...2L, 2017A&A...608A..45L}. Also, it was noticed that the mean albedo of the red 
Centaurs is $\sim 8-12\%$ whereas for the gray ones is lower $\sim 5-6\%$ \citep{2013ApJ...773...22B, 2014A&A...564A..92D}.

Several surveys have been dedicated to the search of small bodies 
\citep[\eg ][]{2008ssbn.book...71P, 2008ssbn.book..573T, 2014AJ....148...55A, 2016arXiv160704895W, 2016ApJ...825L..13S, 2018ApJS..236...18B}. 
However, these surveys are mostly focused on the search of TNOs or NEOs and the discovery of Centaurs is mainly serendipitous. 
Size distributions have been estimated, but they are not debiased \citep{2013ApJ...773...22B}, or used the maximum likelihood technique  \citep{1997Icar..127..494J}, or used Monte Carlo simulations \citep{2000AJ....120.2687S}---these two, however, 
find a similar distribution with $\alpha \sim 0.6$.
\citet{2014AJ....148...55A} proposed a debiased size distribution, 
using Deep Ecliptic Survey data, and inferred that the best fit for Centaurs with absolute magnitudes 
$7.5<H<11$ is a power law with $\alpha$=0.42$\pm$0.02 suggesting a knee in the size distribution around $H=7.2$~mag. Similarly, using the OSSOS simulator, \citet{2018AJ....155..197L} confirmed that a  single slope for the size distribution is not able to match the observations, and thus a break is needed at smaller sizes. 
Such a break is also required to explain the size distribution of the TNOs \citep{2004AJ....128.1364B, 2013ApJ...764L...2S}. 


\startlongtable
\begin{longrotatetable}
\setlength{\tabcolsep}{3pt} 
\renewcommand{\arraystretch}{0.6} 
\begin{deluxetable}{lccccccc}
\tabletypesize{\footnotesize}
\rotate
\tablecaption{\label{Tab:LCs} B$-$R colors of ``strict'' Centaurs from TRC survey$^*$ and their albedos, lightcurves studies, 
and detected ices from other works.}
\tablewidth{0pt}
\tablehead{
Object & B-R & Albedo & Single peak P. [h] & Double peak P. [h]& $\Delta m$ [mag] & Ices? & References  
}
\startdata
29P Schwassmann-Wachmann 1 & - &   4.61$_{-1.90}^{+5.22}$  & -& - & - &-  &  S08\\
(2060) 1977~UB Chiron & - &   7.57$_{-0.87}^{+1.03}$ & -& 5.9180$\pm$0.0001 &  0.088$\pm$0.003 & Water &  B89, O15, G16, SS08\\
(5145) 1992~AD Pholus &1.97$\pm$0.11& 15.5$_{-4.9}^{+7.6}$ & - &  9.98 &  0.15/0.60 & Methanol &  B92,H92,F01,T05, D14 \\
  &-  &-&  - &  - &  0.15 &  - &  RT99\\
(7066) 1993~HA$_{2}$ Nessus &- &  8.6$_{-3.4}^{+7.5}$ &  - &-&  0.5 &  - &  RT99\\
  &- & - & - &  - &  $<$0.2 &  - &  D98, D14\\
 &-  &- & 5.68$\pm$0.19 &  - &  0.76$\pm$0.08 &  - &  K06\\
 1994~TA & 1.92$\pm$0.10  &-&  - &  - &- &  - &  TR00 \\
 (10370) 1995~DW$_{2}$ Hylonome & 1.16$\pm$0.09  & 5.1$_{-1.7}^{+3.0}$ &  - &  - &- &  - &  TR98, RT99, D14\\
  (8405) 1995~GO Asbolus &  1.22$\pm$0.05 & 5.6$_{-1.5}^{+1.9}$ &- &  8.9351$\pm$0.003 &  0.55 & Water &  D98,K00, D14\\
 &-  & -& - &  - &  0.34 &  - &  RT99\\
 & -  &-&  - &  - &  0.14$\pm$0.10 & -&  R07\\
(10199) 1997~CU$_{26}$ Chariklo &1.25$\pm$0.05 &   5.63$_{-0.65}^{+0.76}$ &  - &  -& $<$0.1 & Water &  D98, TR98, RT99, S08\\
 			&-& 5.59$_{-1.17}^{+1.69}$ &  - &  7.004$\pm$0.036 &  0.11 &  -& F14, G16 \\
(49036) 1998~QM$_{107}$ Pelion &1.25$\pm$0.04  & -& - &  - &- &  - & TR00 \\		
(52872) 1998~SG$_{35}$ Okyrhoe & 1.21$\pm$0.02 & 5.6$_{-1.0}^{+1.2}$ &-  &  16.6 &  0.2 &  Water &  B03, D14, TR03\\
  &- &-& 4.68 or 6.08 & - & 0.07$\pm$0.01 &  - &  T10\\
  (52975) 1998~TF$_{35}$ Cyllarus  & 1.72$\pm$0.12 & 13.9$_{-6.4}^{+15.7}$ &-  &  - & - & - &   D14, TRC03\\
  (29981) 1999~TD$_{10}$  &-   & 4.40$_{-0.96}^{+1.42}$   & - &  - &- &  - & S08 \\	
 (31824) 1999~UG$_{5}$ Elatus & 1.75$\pm$0.05& 4.9$_{-1.6}^{+2.8}$&    13.25 &  - &  0.24 & Water & D14, TR03, G01\\
  & &-&  13.41$\pm$0.04 &  - &  0.102$\pm$0.005 &  - &  B02\\
 (60558) 2000~EC$_{98}$ Echeclus & 1.39$\pm$0.04 & 5.2$_{-0.71}^{+0.70}$&    13.401 &  26.802 &  0.24$\pm$0.06 & - &  R05, D14\\
2000~FZ$_{53}$  &1.17$\pm$0.05  & -& - &  - &- &  - & TRC03 \\	
2000~GM$_{137}$ & -  & 4.3$_{-1.6}^{+2.6}$ & - &  - &- &  - & D14 \\	
 (54598) 2000~QC$_{243}$ Bienor & 1.15$\pm$0.08& 4.3$_{-1.2}^{+1.6}$ &   4.57$\pm$0.02 &  9.14$\pm$0.04 &  0.08 to 0.75 &Methanol &  O03, F17, D14\\
 & &-&     - &  9.17 &  0.34$\pm$0.08 &-&  R07\\
 & &-&    (4.723 or 4.594)$\pm$0.001 &  - &  0.38$\pm$0.02 & - &  K06\\
   (63252) 2001~BL$_{41}$  & 1.21$\pm$0.03 & 4.3$_{-1.4}^{+2.8}$ &-  &  - & - & - &   D09, TRC03\\
   (88269) 2001~KF$_{77}$  & 1.81$\pm$0.04 & - &-  &  - & - & - &   TRC03\\
  (32532) 2001~PT$_{13}$ Thereus & 1.18$\pm$0.01& 8.3$_{-1.3}^{+1.6}$ &    4.1546$\pm$0.0005 &  8.3091$\pm$0.0001 &  0.16$\pm$0.02 &Water&  O03, D14 \\
&   - &-&  - &  8.34 &  0.16 &  - &  FD03\\
 & - &-&  - &  8.34 &  0.34$\pm$0.08 &  - &  R07\\
 &  &-&  - &  8.4 &  0.15 &- &  F01\\
    (119315) 2001~SQ$_{73}$  & 1.13$\pm$0.02 & 4.8$_{-1.8}^{+3.0}$ &-  &  - & - & - &   D14, TRC03\\
   (148975) 2001~XA$_{255}$ &-&-&   - & - &  $\sim$0.13 & -& T13 \\
           & -  &-&  $>$7 &  $>$14 &  $\sim$0.2 &- &  H18\\
2001~XZ$_{255}$  & 1.91$\pm$0.07 &- &-  &  - & - & - &   TRC03 \\   
     (42355) 2002~CR$_{46}$ Typhon &- & 5.09$_{-0.80}^{+1.24}$ &  9.67 & - &  0.07$\pm$0.01 &  - &  T10\\        
 (55567) 2002~GB$_{10}$ Amycus &1.82$\pm$0.03 & 8.3$_{-1.5}^{+1.6}$ &    9.76 & - &  0.16$\pm$0.01  & Water & TRC03, T10\\
(83982) 2002~GO$_{9}$ Crantor &1.86$\pm$0.02&  11.8$_{-?}^{+7.09}$ & (6.97 or 9.67)$\pm$0.03 &  - &  0.14$\pm$0.04 & Methanol &  O03, TR03, S08\\
    &- &-&  - &  - &  0.34 &  - &  RT99\\
   (95626) 2002~GZ$_{32}$&1.03$\pm$0.04&-&    - &  5.80$\pm$0.03 &  0.15$\pm$0.03 &  - &  D08\\
   (250112) 2002~KY$_{14}$  & 1.75$\pm$0.02 & - &-  &  - & - & - &  TRC16\\
      (73480) 2002~PN$_{34}$&- &4.25$_{-0.65}^{+0.83}$&    - &  - &- &  - &  S08 \\
   2002~PQ$_{152}$  & 1.85$\pm$0.05 & - &-  &  - & - & - &   TRC16\\
  (523597) 2002~QX$_{47}$  & 1.08$\pm$0.04 & - &-  &  - & - & - &  TRC16\\
(119976) 2002~VR$_{130}$  &- & 9.3$_{-3.6}^{+6.6}$ &-  &  - & - & - &   D14\\
   (120061) 2003~CO$_{1}$ &- & 4.9$_{-0.6}^{+0.5}$ &    3.53/4.13/4.99$\pm$0.01/6.30 &  - &  0.10$\pm$0.05 &- &  O06, D14\\
  &- &-&  4.51 &  - &  0.06$\pm$0.01 &  - &  T10\\
       (65489) 2003~FX$_{128}$ Ceto &- & 7.67$_{-1.10}^{+1.38}$ &  - &4.43$\pm$0.03 &  0.13$\pm$0.02 &  - &  D08, S08 \\  
(136204) 2003~WL$_{7}$&1.23$\pm$0.04& 5.3$\pm$1.0 &  8.24 &  - &  0.04$\pm$0.01 & - &  T10, TRC16, D14\\
   2004~QQ$_{26}$  & - & 4.4$_{-1.4}^{+3.9}$ &-  &  - & - & - &   D14\\
  (447178) 2005~RO$_{43}$  & 1.24$\pm$0.03 & 5.6$_{-2.1}^{+3.6}$ &-  &  - & - & - &   D14, TRC16\\
(145486) 2005~UJ$_{438}$& 1.64$\pm$0.04 & 25.6$_{-7.6}^{+9.7}$  &   8.32 &  - &  0.11$\pm$0.01 &  -&  T10, TRC16, D14\\
(248835) 2006~SX$_{368}$& 1.22$\pm$0.02 & 5.2$_{-0.6}^{+0.7}$  &  - &  - &  -&  -&    TRC16, D14\\
(309139) 2006~XQ$_{51}$& 1.15$\pm$0.03 & -  &  - &  - &  -&  -&  TRC16 \\
(341275) 2007~RG$_{283}$& 1.26$\pm$0.03 & - &  - &  - &  -&  -&  TRC16 \\
2007~RH$_{283}$& 1.15$\pm$0.03 & - &  - &  - &  -&  -&  TRC16 \\
2007~TK$_{422}$& 1.22$\pm$0.04 & - &  - &  - &  -&  -&  TRC16 \\
(25012) 2007~UL$_{126}$ & 1.75$\pm$0.02 &   5.7$_{-0.7}^{+1.1}$&    3.56 or 4.2 &  7.12 or 8.4 &  0.11$\pm$0.01 & -&  T10\\
2007~UM$_{126}$& 1.13$\pm$0.03 & - &  - &  - &  -&  Water &  TRC16 \\
2007~VH$_{305}$& 1.18$\pm$0.02 & - &  - &  - &  -&  -&  TRC16 \\
(281371) 2008~FC$_{76}$& 1.60$\pm$0.03 & 6.7$_{-1.1}^{+1.7}$ &   - &  - & $\sim$0.1 & Methanol &   T13, TRC16, D14 \\
  &- &-&  5.93$\pm$0.05 &  11.86$\pm$0.05 &  0.04$\pm$0.01 &  -& H18\\
(315898) 2008~QD$_{4}$&1.20$\pm$0.02&-&  - &  - & $\sim$0.09 &  -&   T13, TRC16 \\
      & -   &-&  $>$7 &  $>$14 &  $\sim$0.15 &  - &  H18\\
(309737) 2008~SJ$_{238}$& 1.60$\pm$0.02 & - &  - &  - &  -&  -&  TRC16 \\   
(309741) 2008~UZ$_{6}$& 1.52$\pm$0.04 & - &  - &  - &  -&  -&  TRC16 \\
(342842) 2008~YB$_{3}$ & 1.23$\pm$0.02& - &  - &  - & $\sim$0.18 &- &   T13, TRC16 \\
(346889) 2009~QV$_{38}$ Rhiphonos & 1.37$\pm$0.02 & - &  - &  - &  -&  -&  TRC16 \\
 (349933) 2009~YF$_{7}$& 1.18$\pm$0.03 & - &  - &  - &  -&  -&  TRC16 \\     
 2010~BK$_{118}$&-&- &   - &  - & $\sim$0.15 &- &   T13 \\
 (382004) 2010~RM$_{64}$ &1.56$\pm$0.02 & - &  - &  - & - &- &  TRC16 \\
2010~TH &1.18$\pm$0.03 & - &  - &  - & - &- &  TRC16 \\
2010~TY$_{53}$&-&-&   - &- & $<$0.14 &- &  BS13 \\ 
     (471339) 2011~ON$_{45}$ &1.81$\pm$0.04 & - &  - &  - & - &- &  TRC16 \\
  (449097) 2012~UT$_{68}$ &1.68$\pm$0.03 & - &  - &  - & - &- &  TRC16 \\
   (463368) 2012~VU$_{85}$ &1.70$\pm$0.07 & - &  - &  - & - &- &  TRC16 \\
    (523676) 2013~UL$_{10}$ &1.17$\pm$0.03 & - &  - &  - & - &- &  TRC16 \\
(459865)  2013~XZ$_{8}$ &1.17$\pm$0.03 & - &  - &  - & - &- &  TRC16 \\
   (459971) 2014~ON$_{6}$ &1.55$\pm$0.03 & - &  - &  - & - &- &  TRC16 \\
    \hline
    \enddata
   \end{deluxetable}   
\footnotesize
  $^*$ Tegler, Romanishin, and Consolmagno's survey: {\url{http://www.physics.nau.edu/~tegler/research/survey.htm}} \\
  References: 
B89: \citet{1989Icar...77..223B}; 
B92: \citet{1992Icar..100..288B}; 
B02: \citet{2002PASP..114.1309B}; 
B03: \citet{2003Icar..166..195B}; 
BS13: \citet{2013AJ....145..124B}; 
D98: \citet{1998Icar..134..213D}; 
D08: \citet{2008A&A...490..829D}; 
D09: \citet{2009A&A...505.1283D}; 
D14: \citet{2014A&A...564A..92D}; 
F01: \citet{2001Icar..152..238F}; 
F14: \citet{2014A&A...568L..11F}; 
F17: \citet{2017Icar..295...34F}; 
FD03: \citet{2003Icar..164..418F}; 
G01: \citet{2001A&A...371L...1G}; 
G16: \citet{2016Ap&SS.361..212G}; 
H92: \citet{1992LIACo..30..203H}; 
H18: \citet{2018MNRAS.474.2536H}; 
K00: \citet{2000ApJ...542L.155K}; 
K06: \citet{2006PhDT.........5K}; 
O03: \citet{2003A&A...407.1149O};  
O06: \citet{2006A&A...447.1131O}; 
O15: \citet{2015A&A...576A..18O}; 
R05: \citet{2005Icar..176..478R}; 
R07: \citet{2007AJ....133...26R}; 
RT99: \citet{1999Natur.398..129R}; 
S08: \citet{2008ssbn.book..161S}; 
SJ02: \citet{2002AJ....124.1757S}; 
T05: \citet{2005Icar..175..390T}; 
T10: \citet{2010A&A...522A..93T}; 
T13: \citet{2013PhDT.........T}; 
TR00: \cite{2000Natur.407..979T}; 
TR03: \citet{2003Icar..161..181T}; 
TRC03: \citet{2003ApJ...599L..49T}; 
TRC16: \citet{2016AJ....152..210T}.
     \end{longrotatetable}
 

\subsection{Moons and rings}
\label{sec:moons_rings}

Binary and multiple systems have been found across many minor body populations. The Centaurs are no exception with two known binary systems: Ceto/Phorcys---formerly, (65489) 2003~FX$_{128}$---and Typhon/Echidna---formerly, (42355) 2002~CR$_{46}$. 
In the case of Ceto/Phorcys, the components have a radii of 87 and 66~km and are separated by about 1800~km 
and the system has a density of $\sim 1.4$~g cm$^{-3}$\citep{2007Icar..191..286G}. 
Typhon/Echidna has a similar semi-major axis of $\sim$1600~km and the radii of the 
components are 76 and 42~km, with a density of only 0.44~g cm$^{-3}$ \citep{2008Icar..197..260G}. 
Despite finding many binary systems among TNOs, only two were found among Centaurs and there are 
no evidence of contact binary Centaurs so far. But, it is important to keep in mind that the Centaurs have not 
been the topic of any large-scale survey for lightcurve studies. 

Recently, to everyone's surprise, thanks to stellar occultations, a ring system was discovered around the 
largest Centaur: Chariklo \citep{2014Natur.508...72B}. Then, 
\citet{2015A&A...576A..18O, 2015Icar..252..271R,FernandezValenzuela2017} evaluated the possibility of ring systems around Chiron and Bienor. 
The formation and stability of ring systems around objects in the size-scale of $\sim 100$ km is still a mystery. Several hypothesis have been considered to explain the formation of these rings: debris disk from a rotational fission, debris disks due to the disruption of a satellite because of impact and/or tidal forces among other options 
\citep{2016ApJ...828L...8H,2016ApJ...821...18P}. Finally, their stability is also an open question \citep{2018NatAs.tmp..166S}. 



\subsection{Cometary activity}
\label{sec:activity}

As already mentioned, Chiron, the first object identified as a Centaur, was found to exhibit magnitude 
variations on short timescales.
That the first Centaur discovered was an active cometary body was no coincidence. As we saw on Section \ref{sec:progenJFC}
the dynamical evidence for a link between Centaurs and JFCs is compelling. 

From the $\sim 290-360$ Centaurs known, depending on the classification, 29 have shown evidence of activity, 
\ie $\sim 8-9\%$. In addition to the dynamical evidence, then, the Centaur population has the largest fraction of objects with confirmed activity relative to other populations, besides those of actual comet populations, evidently. 
Neither main-belt comets (MBCs)---a.k.a. active asteroids (AAs)---nor NEOs approach the fraction of active bodies embedded within their dynamically defined populations that the Centaurs do. 
Many of the unique attributes of the Centaur population may be linked with their cometary activity and 
behavior, 
and may suggest that activity among the Centaurs is more common than the fraction observed at a given epoch. 
Several attempts have been made to link the bimodal color distribution with current and past activity 
of Centaurs as we saw on Section \ref{sec:colors}. 
A significant color-based correlation with Centaur albedo has been established 
\citep{2008ssbn.book..161S, 2008ssbn.book..105T, 2013ApJ...773...22B, 2014ApJ...793L...2L}. 
The darker population matching the neutrally-colored populations marks a convergence similar to comet surfaces. 
In fact, all small body populations nearer the Sun are composed entirely of dark and neutrally colored objects. 
A plausible explanation for the lack of red and bright surfaces among Centaurs and JFCs is the destruction
of these very same red surfaces when entering the inner solar system over a lifetime comparable to that of 
Centaurs \citep{2002AJ....123.1039J}, but TNOs show that something more complex is going on.  
As if doubts existed still, the match between Centaur and JFC size distribution slopes 
\citep{2013Icar..226.1138F, 2013ApJ...773...22B} also evidences a probably similar origin or evolution. 

In fact, technically, Chiron was not the first Centaur discovered, rather that distinction falls to a 
more active body, 29P/Schwassmann-Wachmann 1, 
discovered in 1927. The pattern of behavior of active Centaurs varies with each Centaur. 
While Chiron's activity varies over large timescales \citep[][and references therein]{2002Icar..160...44D} Echeclus' outburst are shorter lived, but more periodic \citep{2006CBET..563....1C, 2011IAUC.9213....2J, 2019AJ....157...88S}, while 29P's activity is constant, though punctuated with outbursts. 
As suggested in \cite{2009AJ....137.4296J}, the identified active Centaurs tend to have lower 
perihelion (see Figure \ref{fig:a_e}), though notable exceptions, like Chiron, exist. 
In fact, \cite{2018P&SS..158....6F} found that active and inactive Centaurs have experienced different dynamical evolutions. 
They suggest that inactive Centaurs deactivated because of a larger permanence far from the Sun while a recent drastic drop in perihelion distance could be responsible of renew the  activity in the active ones. 
The brightness variations of the other 26 are lower but morphological features are still identifiable \citep{2009AJ....137.4296J}.
Brightness variations and extended morphology are not the sole indicators of possible activity. Larger bodies may exhibit temporary ring structures and dust grain atmospheres as we have seen over Chariklo and Chiron.
No outburst has yet been detected on Chariklo but, as mentioned on Section \ref{sec:moons_rings}, such kind 
of events may both be the origin and eventual demise of the rings. 
The structure surrounding Chiron could also be a manifestation of, or evolve from, a ballistic dust atmosphere \citep{1990AJ....100.1323M}. 

The activity in Centaurs tends to be recurrent but, unlike standard cometary behavior, Echeclus, 29P, and Chiron all exhibit outbursts and coma decoupled from the Centaur's perihelion. The mechanism that triggers the sublimation is not clear. 
It may be that localized regions of 
surface ice are directly exposed through a combination of seasonal and diurnal variations on such bodies, rather than gases being released from the sub-surface, or that volatiles may be driven from water ice via the amorphous-crystalline transition as the thermal wave reaches localized sub-surface water ice reservoirs \citep[see][and Section \ref{sec:internal}]{1992ApJ...388..196P, 2009AJ....137.4296J}.  
Dynamical models show that excursions of Centaurs over the inner solar system and subsequent returns to Centaur orbits are common \citep[\eg ][]{2004MNRAS.354..798H}. Deriving any notion of how volatiles, or hyper-volatiles, are distributed on Centaurs must accommodate the perseverance of a fraction of the ices throughout such terrestrial-planet-region excursions, possibly in sub-surface reservoirs that may later be exposed.

\subsection{Interiors}
\label{sec:internal}

The internal structure of Centaurs is, in theory, the result of processes that occurred at three different stages of their history \citep[see][for reviews]{2008SSRv..138..147P, 2008ssbn.book..243C, 2009M&PS...44.1905S, 2015SSRv..197..271G}. First, the internal structure may reflect the formation of Centaurs themselves, a process which is not fully constrained yet. Some Centaurs could be primordial aggregates of planetesimals relatively unaltered by collisions \citep{2014Icar..231..168B}, while some others could be the outcomes of disruptive collision of larger bodies. 
Indeed, \citet{1996A&A...310..999S} and \citet{1997Icar..125...50D} suggested that JFCs---therefore Centaurs too---, could be fragments from parent bodies with initial sizes up to $\sim$100~km. The ``outbursty'' activity of Echeclus could be interpreted as being due to internal heterogeneities, possibly reflecting its rubble-pile nature \citep[either primordial or collisional, see][]{2016MNRAS.462S.432R}. 
In fact, \citet{2010Icar..209..753R} showed that internal heterogeneities could potentially be detected through the activity pattern of a comet, 
where erratic behaviors would be induced by blocks of varying thermo-physical properties randomly distributed inside the nucleus. 

The second stage of processing is related to radiogenic heating, especially their early heating. The effectiveness of ${}^{26}$Al in heating the interior of an icy object strongly depends on the time delay between its formation and the formation of calcium--aluminium--rich inclusions, CAIs \citep[see][]{1995Icar..117..420P, 2008SSRv..138..147P}, which remains unknown. \citet{1987ApJ...319..993P} found that early radiogenic heating by ${}^{26}$Al could lead to the complete crystallization of amorphous ice inside comets and Centaurs, and even melting of large nuclei. Therefore, the early thermal evolution of Centaurs could have led to 
phase transitions and chemical differentiation occurring mostly in the core, while an outer layer might remain pristine. 
We should stress that the internal structure inherited from the first two stages remain speculative: the effects of accretion could be inferred from size, shape or within the activity pattern, but most observations cannot provide us with this information. Early heating should have mostly affected the core of icy objects, which we cannot access from remote observations of their surface. In addition, the third stage, dominated by cometary activity, could erase many of the effects of the two previous stages.

Whatever stratigraphy is produced after these early stages of formation and evolution, the internal structure of Centaurs should be further modified by cometary activity.
Water ice is thermodynamically stable in the giant planet region, \ie it does not sublimate, hence it cannot drive the activity of Centaurs. 
\cite{2009AJ....137.4296J} suggested that crystalline--amorphous water ice transition, releasing trapped volatiles, could be the source of Centaurs activity. \cite{2012AJ....144...97G} studied this scenario concluding that: 
i) crystallization at the surface of Centaurs can be triggered at heliocentric distances as large as 16 au; 
ii) cometary activity due to crystallization could be triggered up to $\sim10-12$ au, which is broadly consistent with the observations of active Centaurs; 
iii) as the crystallization front propagates from the surface toward the interior of Centaurs, cometary activity should stop in a relatively short period of time: outgassing is typically sustained for hundreds to thousands of years. 
The simple fact that active Centaurs do exist shows that a chemical differentiation, due to thermo-physical processing is happening in the giant planet region. 


\section {Discussion}
\label{sec:discussion}

If Centaurs were discovered later than TNOs they probably would not have been named as a class and would be 
just seen as low perihelion SDOs. Yet, they do not cease to surprise and puzzle us  
revealing themselves to be as rich as TNOs alone.  

Identifying the important large-scale processes in the outer Solar System is of the utmost importance to understand 
how the entire Solar System has formed and how it has evolved. Centaurs are 
an essential class of minor bodies to help us do so. No longer simply because they are closer, and therefore brighter, 
than TNOs, allowing for more detailed observational studies, but also because they exhibit peculiar and distinct behaviors that 
seem to be at the core of the deep physical and chemical changes that affect icy bodies when entering the inner Solar System. 

The condition of transition phase between TNOs and JFCs attributed to Centaurs seems unquestionable. However,  
if dynamical models predict that orbital inclinations are not preserved among Centaurs, how come 
red Centaurs have a smaller orbital inclination angle distribution than the neutral ones? Are dynamical models that 
reliable? If Centaurs are scattered inwards from the trans-Neptunian region through interaction with Neptune, how can 
they possibly have a different collisional history than TNOs? Indeed puzzling, but only with more data we can make it indisputable. 
There are no red surfaces among JFCs, therefore red Centaurs must change their colors when becoming JFCs. 
Is it not a remarkable coincidence that neutral Centaurs, that should have formed beyond Jupiter in the primordial disk, 
do not change their colors when becoming JFCs? Too remarkable to leave us comfortable. 
Models for scenarios of distinct compositions over the primordial 
planetesimal disk of icy bodies seem increasingly more sound and refined, 
and yet the evidence for the two distinct 
color groups of Centaurs has faded. Is it not a paradox? 
Bearing in mind the principle of explosion, \ie from a false premise anything follows, models, however elegant they may be, need to be evidence-based. It is through observational data they can be 
demonstrated or refuted. Paradoxically, about 20 years ago we had a seemingly elegant model capable of explaining a 
continuous color distribution and colors started to appear noncontinuous, whereas today colors might be continuous and we 
have an apparently elegant model capable of explaining their discontinuity. 
Confirming or refuting is imperative. 
Consensus is not a demonstration.


\acknowledgments

{\it Acknowledgments:} CITEUC is funded by National Funds through FCT -- Foundation for Science and Technology (project: UID/Multi/00611/2013) and FEDER - European Regional Development Fund through COMPETE 2020 – Operational Programme Competitiveness and Internationalization (project: POCI-01-0145-FEDER-006922). Audrey Thirouin is partly funded by Lowell Observatory funds and by the National Science Foundation (NSF), grant number AST-1734484. 

 


\end{document}